\documentclass[prl,twocolumn,superscriptaddress,nobibnotes]{revtex4}

\usepackage{graphicx}
\usepackage{dcolumn}
\usepackage{bm}
\usepackage{times,mathptmx}
\usepackage{color}
\usepackage{multirow}
\usepackage{enumitem}
\usepackage{chngpage}
\usepackage{amsmath}

\begin{document}


\title{\bf Multicolor Bound Soliton Molecule}

\author{Rui Luo}
\thanks{These authors contributed equally to this work.}
\affiliation{Institute of Optics, University of Rochester, Rochester, NY 14627}
\author{Hanxiao Liang}
\thanks{These authors contributed equally to this work.}
\affiliation{Department of Electrical and Computer Engineering, University of Rochester, Rochester, NY 14627}
\author{Qiang Lin}
\email{qiang.lin@rochester.edu}
\affiliation{Institute of Optics, University of Rochester, Rochester, NY 14627}
\affiliation{Department of Electrical and Computer Engineering, University of Rochester, Rochester, NY 14627}



\begin{abstract}
We show a new class of bound soliton molecule that exists in a parametrically driven nonlinear optical cavity with appropriate dispersion characteristics. The composed solitons exhibit distinctive colors but coincide in time and share a common phase, bound together via strong inter-soliton four-wave mixing and Cherenkov radiation. The multicolor bound soliton molecule shows intriguing spectral locking characteristics and remarkable capability of spectrum management to tailor soliton frequencies, which may open up a great avenue towards versatile generation and manipulation of multi-octave spanning phase-locked Kerr frequency combs, with great potential for applications in frequency metrology, optical frequency synthesis, and spectroscopy.

\end{abstract}

\maketitle

Optical solitons represent a fascinating manifestation of nonlinear optical phenomena in nature \cite{KivsharBook}. In an optical cavity, a balanced interaction among Kerr nonlinearity, group-velocity dispersion (GVD), optical gain, and loss results in temporal dissipative solitary waves that underlie a variety of passive mode locking \cite{Grelu12}. With a particle-like nature, dissipative solitons exhibit profound nonlinear optical dynamics that has been extensively explored in the past decades \cite{Grelu12, Lugiato87, Wabnitz92, Akhmediev97, Barland02, Mitschke05, Mitschke08, Lederer09, Coen10}. One intriguing example is the formation of soliton molecules \cite{Grelu12, Akhmediev97} in which multiple identical solitons are bound together with discrete values of phases and temporal separations \cite{Grelu12, Akhmediev97, Mitschke05, Mitschke08, Lederer09}. Recently, temporal dissipative solitons become relevant in the context of optical Kerr frequency comb generation in high-Q microresonators \cite{Kippenberg11_3}, where the soliton formation was shown to be a major mechanism responsible for the phase locking of Kerr frequency comb \cite{Coen13, Gaeta13_2, Kippenberg14, Coen10}. The extremely highly nonlinear nature of the systems results in complicated interplay between the underlying four-wave mixing (FWM) process and the device dispersion, whose exact nature is currently under intensive investigation \cite{Matsko11, Kippenberg11, Vahala12, Gaeta11_2, Dudley12, Matsko12_2, Chembo13, Coen13_2, Gaeta13, Wabnitz13, Tang13, Zhang13, Diddams14, Gelens14, Gaeta14, Skryabin14, Zeng14, Gelens14_2, Kippenberg_arXiv, Wong15, Weiner15}.

Here we show a new class of bound soliton molecule that forms during Kerr comb generation. The solitons exhibit distinctive colors, coincide in time, and share a common phase, bound together via strong inter-soliton FWM and inter-soliton Cherenkov radiation, in opposite to conventional soliton molecules that consist of a single color while separated in time \cite{Akhmediev97}. As we will show below, the spectral locations of multicolor solitons can be tailored to far separate frequencies without sacrificing their amplitudes, in strong contrast to current Kerr comb generation where the soliton spectrum is primarily located around the pump frequency with a bandwidth dependent on the GVD and intracavity pump power \cite{Coen13, Coen13_2, Kippenberg14}. The generation of such a multicolor bound soliton molecule offers an elegant solution to produce multi-octave ultra-broadband phase locking of Kerr frequency combs that are essential for broad applications such as frequency metrology, precision spectroscopy, and photonic signal processing \cite{Hansch02, Newbury11}.

The dynamics of an optical wave inside a parametrically driven cavity is described by the generalized Lugiato-Lefever equation \cite{Lugiato87, Wabnitz92, Coen13, Note3}
\begin{eqnarray}
	t_R \frac{\partial E}{\partial t}  = && (-\frac{\kappa_t}{2}-i\Delta_0) E + \sum_{m=2}^{\infty} {\frac{i^{m+1} \beta_m}{m!} \frac{\partial^m E}{\partial \tau^m} } \nonumber \\
	&& + i \gamma L \left(1 + \frac{i}{\omega_0}  \frac{\partial }{\partial \tau} \right) \left( |E|^2 E \right) + \sqrt{\kappa_e} A_0, \label{LLE}
\end{eqnarray}
where $E(t,\tau)$ is the amplitude of the intracavity field, with $t$ and $\tau$ representing the slow and fast times, respectively. $A_0$ is the input continuous-wave (CW) pump field launched at frequency $\omega_0$, with a laser-cavity detuning of $\Delta_0=M_02\pi-\omega_0t_R$, where $M_0$ is the order of cavity resonance closest to the pump field and  $t_R$ is the roundtrip time. The cavity has a circumference of $L$, and a nonlinear parameter $\gamma = \frac{\omega_{\rm 0} n_2}{c A_{\rm eff}}$, where $n_2$, and $A_{\rm eff}$ are the Kerr nonlinearity coefficient and effective mode area, respectively, and $c$ is the velocity of light in vacuum \cite{Coen13, Wabnitz92}. To describe the optical wave with a broadband spectrum, Eq.~(\ref{LLE}) includes the self-steepening effect and high-order dispersion where $\beta_m$ denotes the $m$-th order dispersion coefficient at the pump mode. $\kappa_t=\kappa_0+\kappa_e$ represents the total power loss per round trip, where $\kappa_0$ and $\kappa_e$ are the intrinsic loss and pump power coupling coefficients, respectively.

We search for a steady-state solution of multicolor bound solitons in the following form
\begin{equation}
	E(t,\tau) \approx E_0 + \sum_{n=1}^N E_n {\rm sech} ({\tau}/{T_n}) e^{-i (\omega_n-\omega_0) \tau} e^{-i\phi_n}, \label{EField}
\end{equation}
where $E_0$ is the CW field background inside the cavity. $E_n$, $T_n$, and $\phi_n$ are the amplitude, temporal width, and phase, respectively, of the $n$-th soliton at frequency $\omega_n$. When $N=1$, it reduces to the case of single-color solitons which are supported by a constant device GVD \cite{Coen13, Kippenberg14}. The fundamental reason underlying the single-color soliton formation lies in that a constant GVD leads to group index (and thus the mode spacing of the cavity resonances) dependent linearly on the resonance frequency. Only a certain amount of mode spacing mismatch can be compensated by the nonlinear phase modulations of the intracavity field, which leads to a limited soliton spectrum centered around the pump frequency \cite{Coen13_2}.

\begin{figure}[b!]
\includegraphics[width=1.0\columnwidth]{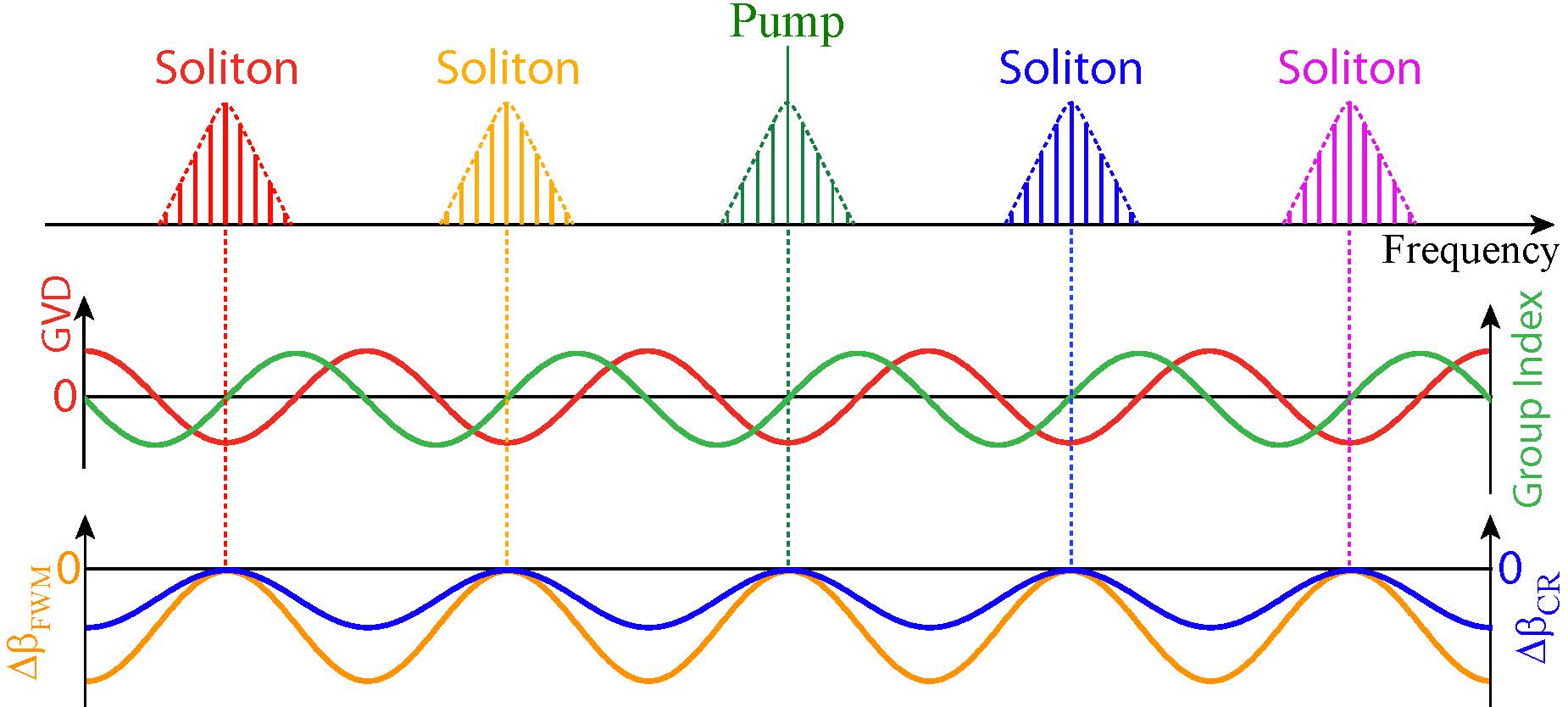}
\caption{\label{Fig1} Schematic of a multicolor bound soliton molecule and the corresponding sinusoidally oscillatory GVD (red), group index (green), phase mismatch for FWM (orange), and phase mismatch for Cherenkov radiation (blue).}
\end{figure}
For a multicolor bound soliton state, the solitons with different colors should have group velocities close enough such that they can remain overlapping in time for strong inter-pulse interaction. As bright solitons generally form in the anomalous dispersion regime, this implies that the device should exhibit a dispersion that is anomalous in multiple spectral regimes where the soliton spectra are located while with group velocities matched with each other. We thus speculate that the optimal condition for a multicolor bound soliton state would be a cavity dispersion oscillatory around zero, as schematically shown in Fig.~\ref{Fig1}. A simple example of oscillatory GVD is a sinusoidal function of frequency given as $\beta_2(\omega) = B \cos(\frac{\omega-\omega_0}{\Omega})$ with $B < 0$, which corresponds to a group index $n_g(\omega)$ and a propagation constant $\beta(\omega)$ given as
\begin{eqnarray}
n_g(\omega) &=& n_g(\omega_0) + B c \Omega \sin(\frac{\omega-\omega_0}{\Omega}), \label{n_g}\\
\beta(\omega) &=& \beta(\omega_0) + \frac{n_g(\omega_0)}{c} (\omega - \omega_0) + 2 B \Omega^2 \sin^2(\frac{\omega-\omega_0}{2\Omega}). \quad \label{beta}
\end{eqnarray}
For the degenerate FWM process $2\omega_0 \rightarrow \omega_s + \omega_i$ responsible for the primary comb generation, such a device dispersion results in a linear phase mismatch of
\begin{eqnarray}
\Delta\beta_{\rm FWM}(\omega_s) &\equiv& \beta(\omega_s) + \beta(\omega_i) - 2\beta(\omega_0) \nonumber\\
&=& 4B \Omega^2\sin^2(\frac{\omega_s - \omega_0}{2\Omega}), \label{kappa}
\end{eqnarray}
which is small for cavity modes around frequencies $\omega_0 + M 2\pi \Omega$ ($M$ is an integer) and can be compensated by the nonlinear phase shifts induced by the pump wave. Consequently, the pump wave would produce a primary frequency comb in these anomalous-dispersion regions with uniform efficiency that would evolve into multiple solitons with different colors but similar amplitudes.

\begin{figure}[t!]
	\includegraphics[width=1.0\columnwidth]{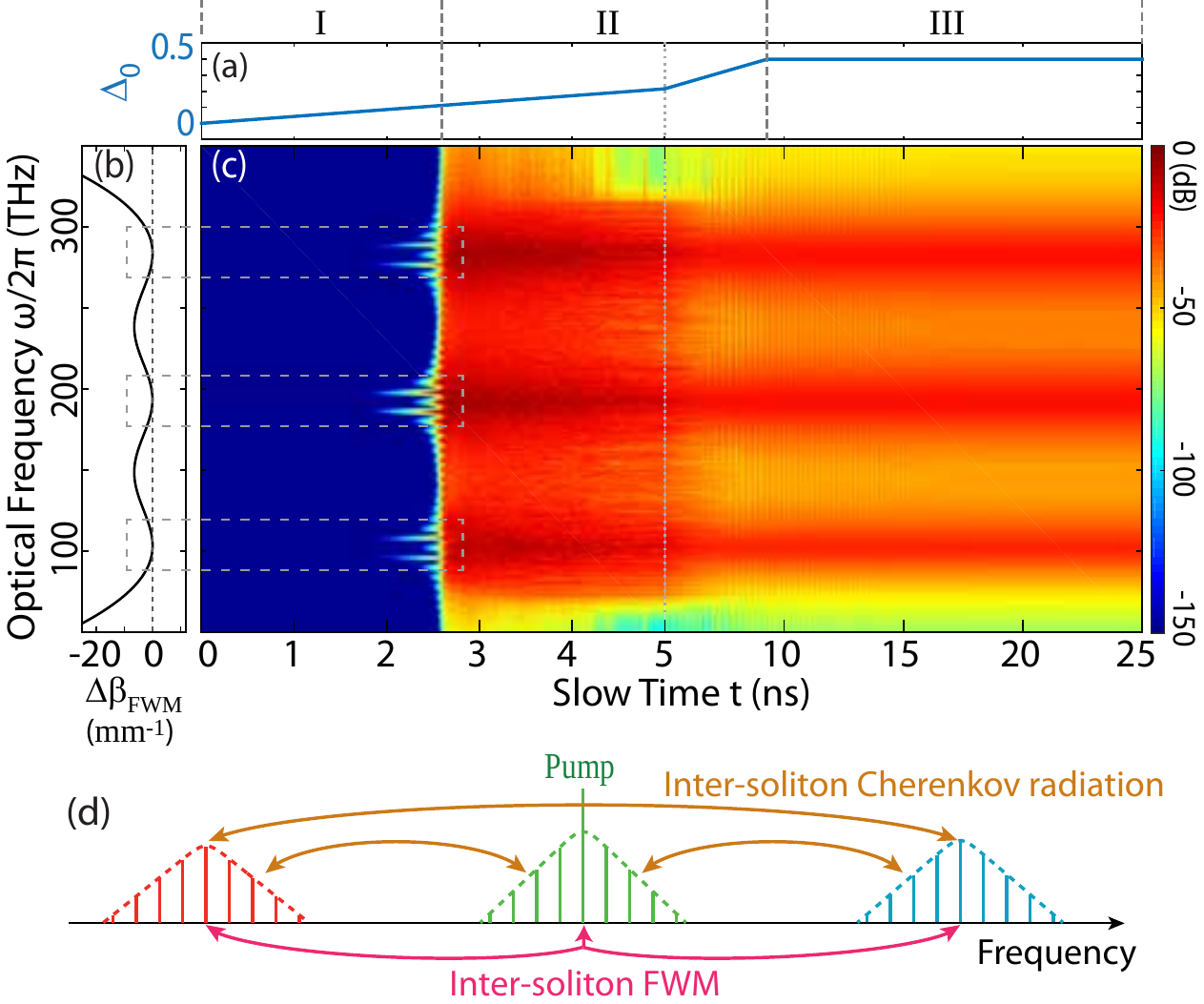}
	\caption{\label{Fig2} Formation dynamics of a three-color bound soliton molecule. (a) Laser-cavity detuning $\Delta_0$ increases linearly from 0 to 0.4 within 9.29~ns and stays constant afterwards. (b) Phase mismatch $\Delta\beta_{\rm FWM}$ for FWM process. (c) Spectral growth of the soliton molecule. The scaling of the time axis changes at $t=5$~ns (the vertical dashed line) to show the formation details of soliton molecule and its stability, which is also responsible for the kink in (b). (d) Schematic of the mode locking mechanism. The device parameters are given in \cite{Note2}. }
\end{figure}
In particular, it is straightforward to show that the phase matching condition is satisfied for FWM among these solitons located around $\omega_j \approx \omega_0 + M_j 2\pi \Omega$ ($j=1-4$), $\omega_1 + \omega_2 \rightarrow \omega_3 + \omega_4$: \cite{Note1}
\begin{eqnarray}
\beta(\omega_3) + \beta(\omega_4) - \beta(\omega_1) - \beta(\omega_2) \approx 0, \label{kappa2}
\end{eqnarray}
which would lead to strong \emph{inter-pulse FWM} among the solitons. Moreover, Eqs.~(\ref{n_g}) and (\ref{beta}) show that these solitons also directly match their phases since \cite{Note1}
\begin{eqnarray}
\Delta \beta_{\rm CR} (\omega_{i},\omega_{j}) &\equiv& \beta(\omega_i) - \beta(\omega_j) - \frac{n_g(\omega_i)}{c}(\omega_i-\omega_j) \approx 0. \quad \label{beta_CR}
\end{eqnarray}
This would result in strong \emph{Cherenkov radiation} between the solitons \cite{Dudley06, Skyriabin10}. As these solitons matches their group indices (Eq.~(\ref{n_g})), they overlap with each other all the time, resulting in significant \emph{inter-soliton FWM} and \emph{Cherenkov radiation} which not only balance soliton energies, but also lead to strong phase locking among the multicolor solitons, eventually forming a bound state of a soliton molecule.

\begin{figure}[t!]
	\includegraphics[width=1.0\columnwidth]{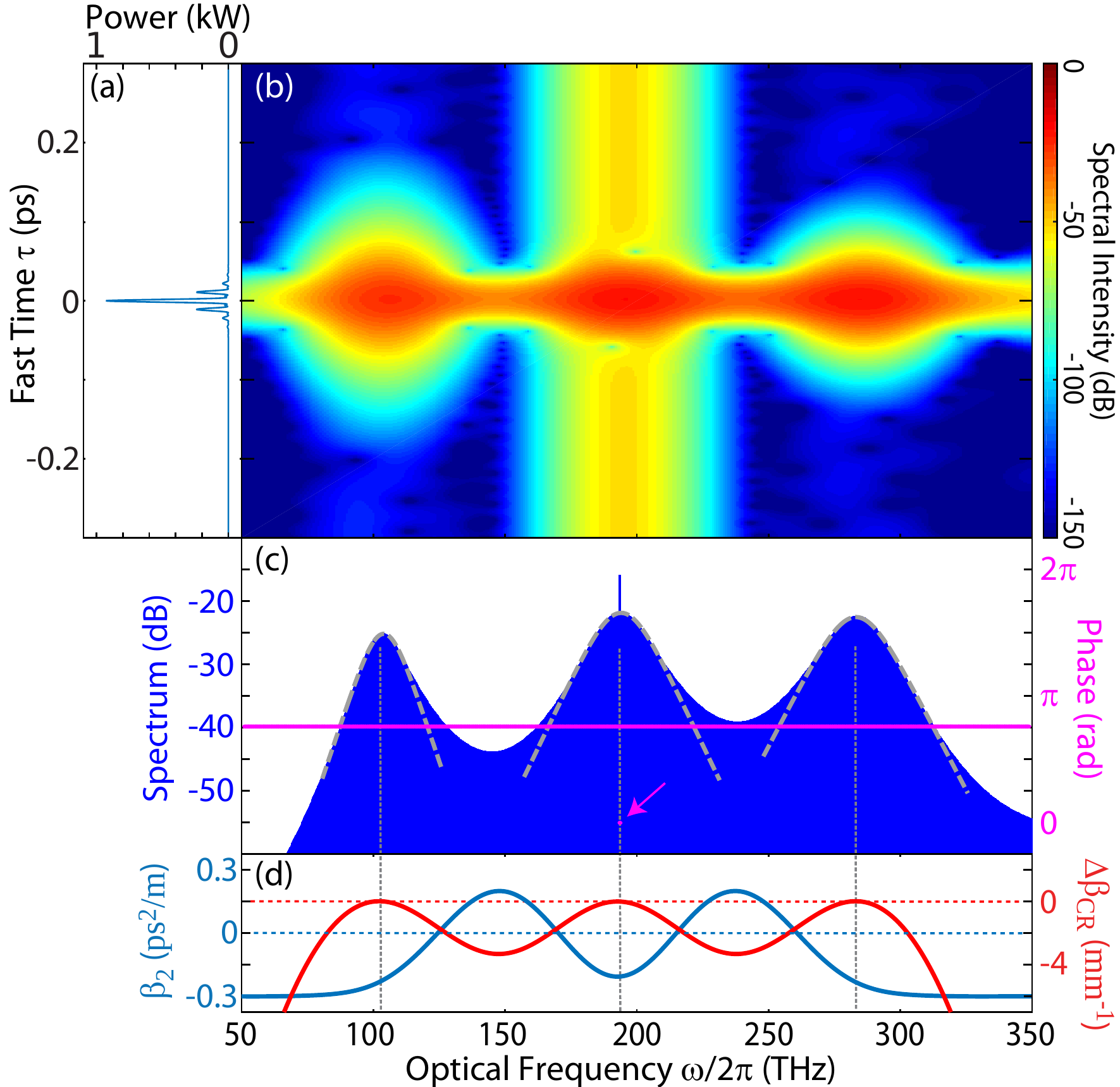}
	\caption{\label{Fig3} Temporal and spectral structure of a three-color soliton molecule. (a) Pulse waveform. (b) Spectrogram. (c) Spectrum (blue) and phase (magenta). The arrow indicates the discrete phase of the pump mode. The dashed curves are hyperbolic-secant fittings of the three distinct spectral components. (d) GVD (blue) and corresponding $\Delta\beta_{\rm CR}(\omega_0, \omega)$ (red), where $\omega_0$ is the pump frequency. }
\end{figure}

To show this concept, we construct an example of a device GVD as shown in the blue curve of Fig.~\ref{Fig3}d, which oscillates sinusoidally between 100 and 300 THz. Such a dispersion curve leads to phase matching, $\Delta \beta_{\rm FWM} \approx 0$, in the spectral regions around 100, 190, and 280 THz for the FWM process (Fig.~\ref{Fig2}b). Consequently, pumping at a cavity mode around 194 THz produces a primary comb in these three regions, as shown in Stage I of Fig.~\ref{Fig2}c. With a red detuning of the pump frequency, the primary comb introduces significant modulational instability that extends the spectrum into a broadband frequency comb (Fig.~\ref{Fig2}c, Stage II). Of particular interest is that, with further red tuning of the pump frequency, the frequency comb evolves into three distinctive spectral components centered around 104, 194, and 284 THz, respectively, that are stable over time (Fig.~\ref{Fig2}c, Stage III).

These three spectral components correspond to three solitons with different colors, as shown clearly in Fig.~\ref{Fig3}. All three solitons are free from chirp, coincide in time (Fig.~\ref{Fig3}b), and exhibit hyperbolic-sechant-like spectral shapes (Fig.~\ref{Fig3}c). As the solitons match their phases, $\Delta \beta_{\rm FWM} \approx 0$ and $\Delta \beta_{\rm CR} \approx 0$, as well as group velocities (Fig.~\ref{Fig3}d and Fig.~\ref{Fig2}b), strong inter-soliton FWM and Cherenkov radiation are expected to occur among them (Fig.~\ref{Fig2}d), which bond the solitons together to form a soliton molecule. Such a bound state is not simply a sum of individual solitons, but is supported by strong inter-soliton interactions. This is evident in the valley regions of the comb spectrum which connect the soliton spectra (Fig.~\ref{Fig3}c), where the spectral amplitudes are considerably higher than the sum of individual solitons. The bound soliton molecule state is also directly reflected in the phase of the solitons. As shown in Fig.~\ref{Fig3}c, a common phase is shared across the entire spectrum that is 0.758$\pi$ different from that of the pump, corresponding to $\phi_n=0.758\pi$ in Eq.~(\ref{EField}). In the time domain, the multicolor soliton molecule manifests as an ultrashort pulse with a full-width of half maximum of 4~fs, as shown in Fig.~\ref{Fig3}a. The sidelobes of the pulse indicate the phase coherence and temporal beating among solitons. It is important to note that, over each round-trip time, there is only one single pulse, the multicolor bound soliton molecule, that cycles inside the cavity.

\begin{figure}[t!]
	\includegraphics[width=1.0\columnwidth]{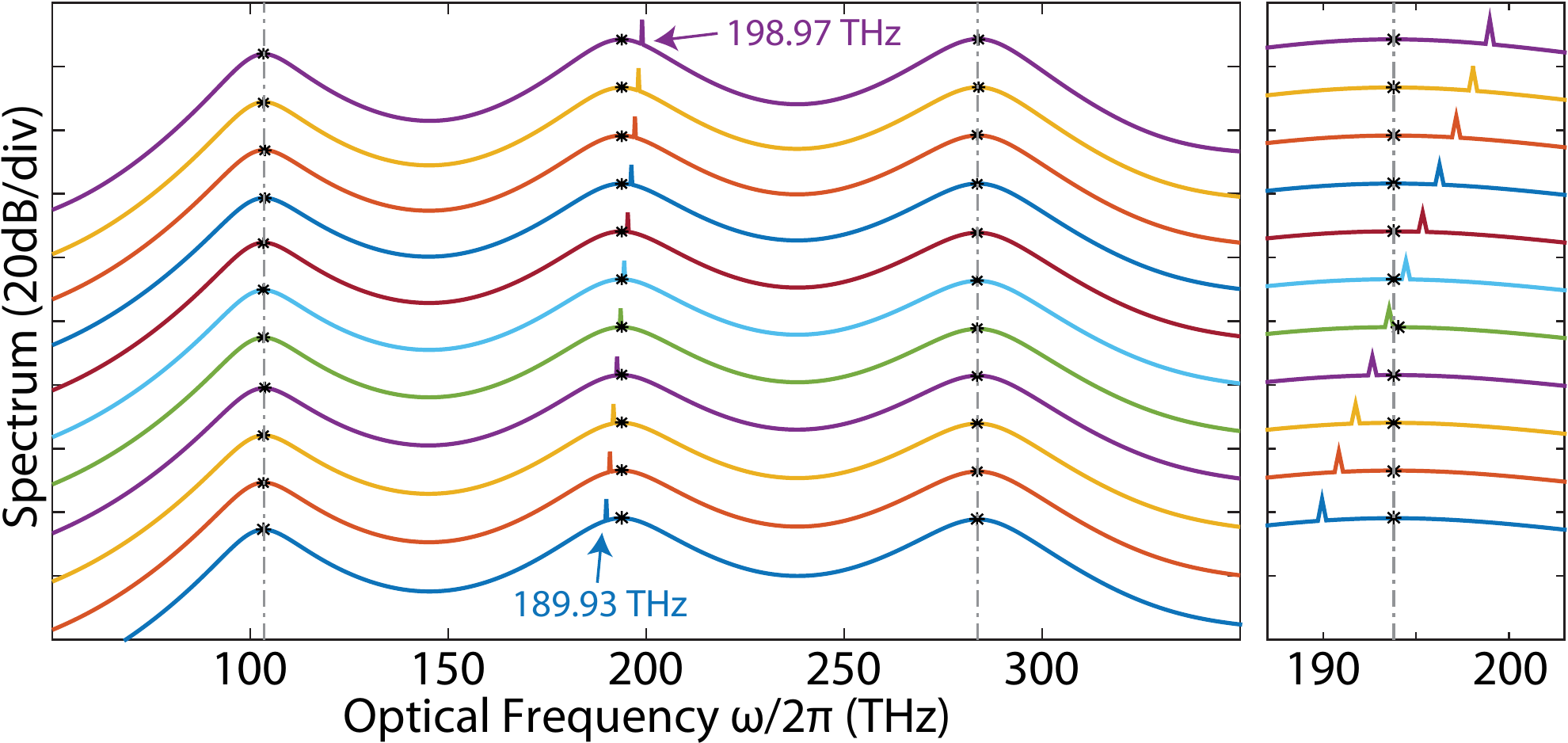}
	\caption{\label{Fig4} Spectra of three-color soliton molecule when the pump frequency varies from 189.93 to 198.97 THz, with a step of $0.904$~THz (corresponding to 4 FSRs). Each spectrum is relatively shifted by 20 dB for better comparison. The stars denote the spectral peaks of the solitons. Right panel: Zoom-in spectra in $187-203$~THz. }
\end{figure}
Figure~\ref{Fig3}c shows that individual solitons inside the soliton molecule exhibit similar spectral amplitudes while their frequencies are separated far apart. Consequently, the whole comb spectrum extends over an broad bandwidth of two octaves, significantly beyond what can be produced by conventional Kerr comb generation mechanisms \cite{Kippenberg11_3, Coen13, Gaeta13_2, Kippenberg14, Vahala12, Matsko11, Kippenberg11, Gaeta11_2, Dudley12, Matsko12_2, Chembo13, Coen13_2, Gaeta13, Wabnitz13, Tang13, Zhang13, Diddams14, Gelens14, Gaeta14, Skryabin14, Gelens14_2, Zeng14, Kippenberg_arXiv, Wong15, Weiner15}. A slight difference among the soliton spectral amplitudes arises from the self-steepening effect: a stronger optical Kerr effect at higher frequency leads to a higher spectral amplitude and a broader spectral width of a soliton component. The uniformity of the soliton energies in a soliton molecule results from the regenerative nature of high-Q cavity which allows multicolor solitons to interact over enough time to balance their energies. This feature is distinctive from conventional resonant wave interaction introduced by simultaneously phase and group velocity matching \cite{Zeng14, Dudley06, Skyriabin10, Trillo12}, where the strength of nonlinear interaction decreases quickly with increased frequency separation. It is also in strong contrast to the Cherenkov radiation in current Kerr frequency combs which only produces narrow band dispersive waves \cite{Coen13_2, Gaeta13, Gelens14, Gaeta14, Skryabin14, Gelens14_2, Zeng14, Kippenberg_arXiv}.

The bound soliton molecule exhibits intriguing spectral locking characteristics. Figure~\ref{Fig4} shows this feature. Nearly identical comb spectra are produced even by tuning the pump frequency considerably from 189.96 to 198.97~THz, across 40 cavity modes. Of particular interest is that the frequencies of the three-color solitons remain intact, as indicated by the dashed lines in Fig.~\ref{Fig4}. The resilience of the soliton molecule to the pump frequency variation is another consequence of the strong inter-soliton nonlinear interaction, which locks the soliton frequencies to where the phase matching of inter-soliton FWM, that of inter-soliton Cherenkov radiation, and group velocity matching are the best satisfied. In this sense, the device GVD functions as a potential array to \emph{spectrally trap} the spectrum of the soliton molecule to the most stable region where the inter-soliton locking is maximized.
\begin{figure}[t!]
	\includegraphics[width=1.0\columnwidth]{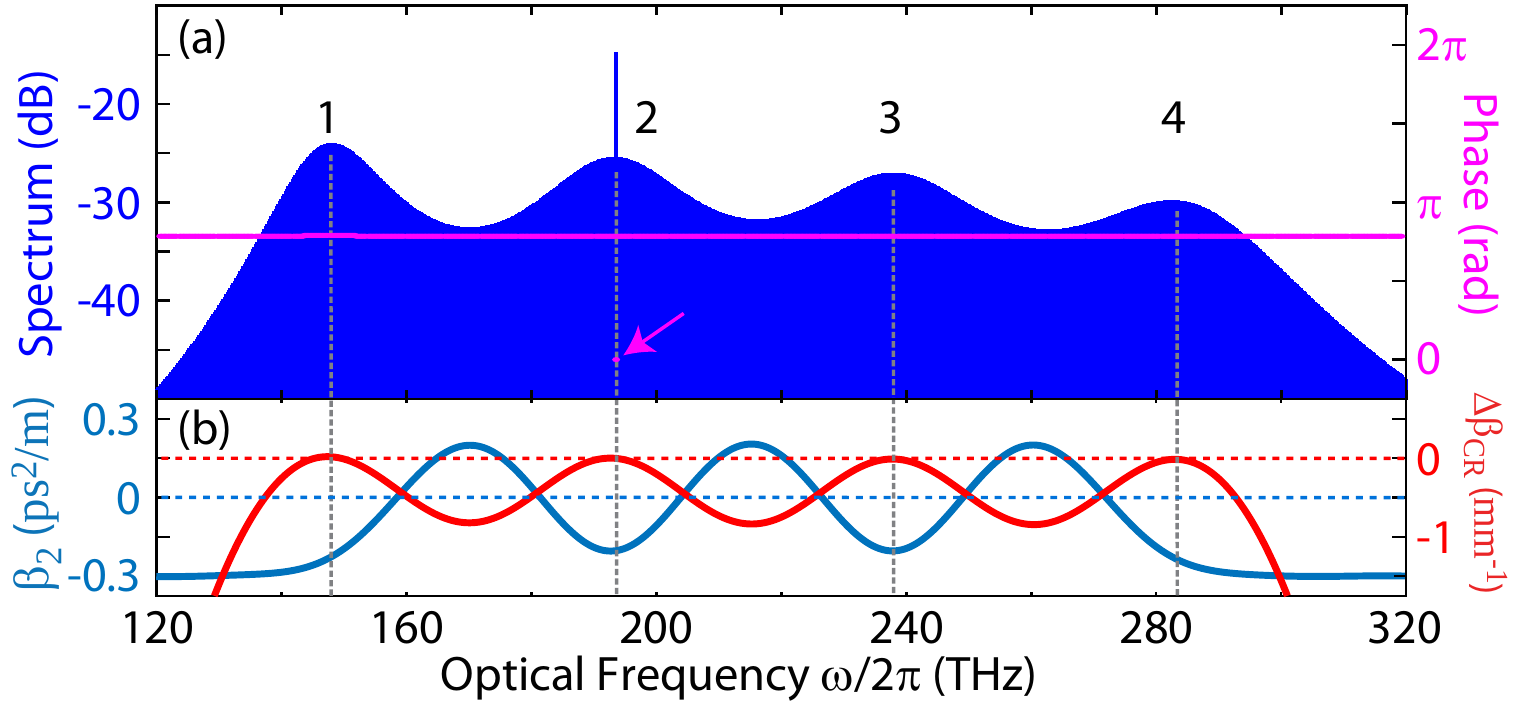}
	\caption{\label{Fig5} (a) Spectrum (blue) and phase (magenta) of a four-color soliton molecule. The arrow indicates the pump phase. (b) GVD (blue) and $\Delta\beta_{\rm CR}(\omega_0, \omega)$ (red). Device parameters are given in \cite{Note2}.}
\end{figure}

\begin{figure}[t!]
	\includegraphics[width=1.0\columnwidth]{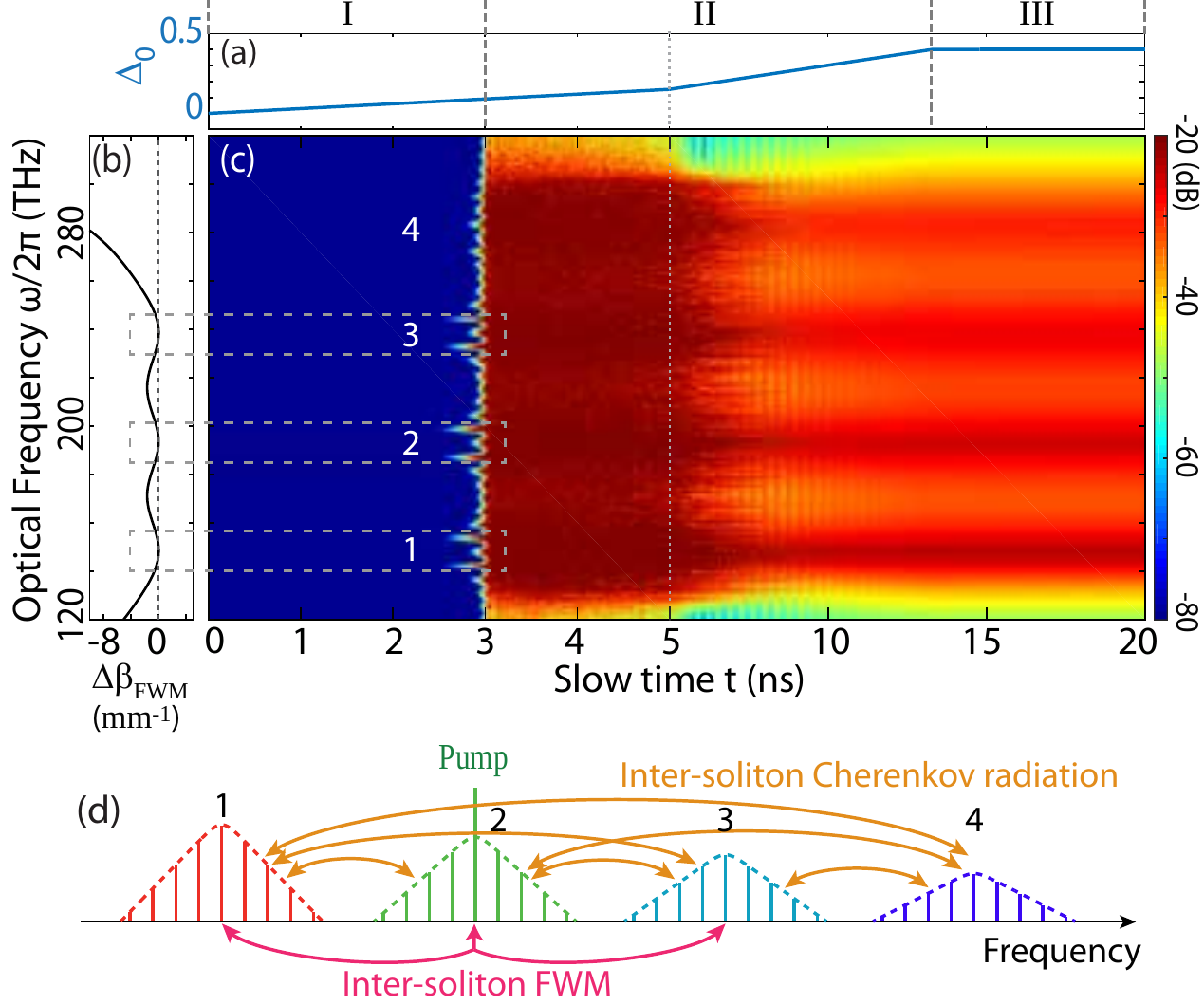}
	\caption{\label{Fig6} Formation dynamics of a four-color bound soliton molecule. The figure structure is the same as Fig.~\ref{Fig2}. In (a), $\Delta_0$ increases linearly from 0 to 0.4 within 13.27~ns and stays constant afterwards. Note again the change of time scaling at $t=5$~ns. }
\end{figure}

The idea of multicolor soliton molecule can be applied to produce soliton molecules that constitute a larger number of soliton components with different colors. Figure \ref{Fig5} and \ref{Fig6} show an example of a four-color soliton molecule. The situation is slightly different from the previous case since the phase matching can only be satisfied among Region 1-3 for the initial FWM process produced by the pump mode (Fig.~\ref{Fig6}b), resulting in primary comb generated in Region 1-3, but not in Region 4 (Fig.~\ref{Fig6}c, Stage I). However, since $\Delta \beta_{\rm CR} (\omega_0, \omega) \approx 0$ among all these four regions (Fig.~\ref{Fig5}b, red), inter-soliton Cherenkov radiation would transfer energy from other three regions to Region 4 (Fig.~\ref{Fig6}c, Stage II), eventually forming a four-color soliton molecule with a common phase of 0.785$\pi$ but tilted soliton amplitudes (Fig.~\ref{Fig5}a). Interestingly, Soliton 1 exhibits a spectral amplitude even higher than Soliton 2 located around the pump mode, which results from the spectral recoil effect \cite{Skyriabin10} that the imbalanced energy transfer towards Soliton 4 produces a spectral recoil towards lower frequency regions. However, as the soliton frequencies are locked due to the \emph{spectral trapping} induced by the device dispersion, the spectral recoil effect manifests here as the amplitude increase of Soliton 1, rather than soliton spectral shifting appearing in conventional systems \cite{Dudley06, Skyriabin10}. 

To date, it remains an open question what is the optimal device dispersion characteristics for generating broadband phase-locked Kerr frequency combs. We hope that our study here provides an answer to this question. Recent advance on dispersion engineering has shown great promise towards this direction \cite{Zhang12, Wang12}. Note that the formation of multicolor soliton molecules is universal as long as the conditions for inter-soliton FWM and Cherenkov radiation are satisfied. The exact values of the amplitude and period of the oscillatory GVD ($B$ and $\Omega$ in Eq.~(\ref{beta})) are not critical. $|B|$ affects primarily the spectral widths of individual solitons, which broaden with decreased $|B|$ and eventually smear together to form a flat comb spectrum. By changing $\Omega$, the soliton frequencies in a soliton molecule can be tailored to desired spectral regions without degrading their amplitudes.

In conclusion, we have demonstrated by numerical modeling a new class of bound soliton molecule generated in a parametrically driven nonlinear cavity. The formation of multicolor bound soliton molecule would have profound impact on Kerr frequency combs. On one hand, it enables producing solitons with frequencies separated by $n$ octaves, ideal for $f$-to-$nf$ interference that is critical for frequency metrology and optical frequency synthesis \cite{Hansch02, Newbury11}. On the other hand, it might function as a universal scheme to produce Kerr comb in the spectral regions that are challenging to access for other approaches. Although we focus here in the context of temporal dissipative solitons, given the space-time duality, the concept of multicolor soliton molecule can be applied to spatial solitons as well. In this case, a certain periodic modulation of spatial frequencies (say, with a gradient-index coupled waveguide array) together with the optical Kerr effect might result in bound soliton molecule that consists of soliton components with intriguing spatial and directional characteristics.

We thank Dr.~Lin Zhang for helpful discussions. This work was supported by the DARPA SCOUT program through grant number W31P4Q-15-1-0007 from AMRDEC.

\end{document}